\documentstyle[psfrag,aps,preprint,epsfig,axodraw]{revtex}

\begin{document}

\tighten

\preprint{TU-730}

\title{Recent Measurements of CP Asymmetries of $B \to \phi K^0$ and
$B \to \eta' K_S$ at $B$-factories Suggest New CP Violation \\
in Left-Handed Squark Mixing}
\author{Motoi Endo, Satoshi Mishima and Masahiro Yamaguchi}
\address{Department of Physics, 
         Tohoku University 
         Sendai, 980-8578, Japan}
\maketitle
\begin{abstract}
  Recent results on the mixing-induced CP asymmetries of $B \to \phi
  K^0$ and $B \to \eta' K_S$ measured at $B$-factories appear to
  indicate discrepancies from the Standard Model
  expectation. Explanation of this possible anomaly is given in the
  context of supersymmetric extension of the Standard Model. It is
  shown that the present data, if the average of two experiments is
  taken, implies additional CP violation appearing in the generation
  mixing of left-handed squarks rather than that of right-handed ones.
  This explanation is argued to disfavor many mediation mechanisms of 
  supersymmetry breaking, including minimal supergravity, because 
  renormalization group effects via Yukawa interaction are unlikely to 
  generate the desired mass mixing.
\end{abstract}

\clearpage

Having increased our confidence on the Kobayashi-Maskawa scheme in the
Standard Model (SM) by precisely measuring $\sin 2 \phi_1/\beta$ from the
$J/\psi K_S$ mode and other charmonium final states in BaBar/Belle
experiments at $B$-factories, the next step forward is to look for hints
of new physics as well as to determine the other CP phases and confirm
the SM in a more convincing way.

Among other things, the measurements of the CP asymmetries of the $b
\to s$ penguin modes are promising, as the SM contribution to $b$ to
$s$ transitions is one-loop suppressed and therefore signatures of new
physics can be quite sizable.  Note that, in the standard model, the
mixing induced CP asymmetries in these processes should be the same as
the one determined by the charmonium final states within roughly 10\%
accuracy, detail depending on the processes.

Possible deviation from the SM was first announced in $B \to \phi K_S$ mode 
in 2002, with subsequent updates of the data in the next year. 
Recently new results on the measurements of the CP asymmetries on the
processes from $b$ to $s \bar q q$ quarks have been 
presented~\cite{Belle,BaBar},
reporting possible anomalies not only in $B
\to \phi K^0$ but also other processes, including $B \to \eta' K_S$.
The Belle experiment has given the new results on the mixing induced
CP asymmetries of $B \to \phi K^0$ and $B \to \eta' K_S$, denoted by
$S_{\phi K^0}$ and $S_{\eta' K_S}$ respectively, as
\begin{eqnarray}
       S_{\phi K^0}(\mbox{Belle})&=&0.06 \pm 0.33 \pm 0.09, 
\\
       S_{\eta' K_S}(\mbox{Belle})&=&0.65 \pm 0.18 \pm 0.04.
\end{eqnarray}
The latest results of the BaBar experiment are, on the other hand,
 \begin{eqnarray}
       S_{\phi K^0}(\mbox{BaBar})&=&0.50 \pm 0.25 ^{+0.07}_{-0.04},
\\
       S_{\eta' K_S}(\mbox{BaBar})&=&0.27 \pm 0.14 \pm 0.03. 
\end{eqnarray}
  If one combines the results from the two experiments,
one obtains~\cite{Ligeti}
\begin{eqnarray}
       S_{\phi K^0}(\mbox{WA})&=& 0.34 \pm 0.21, \label{eq:WA-phiK}
\\
       S_{\eta' K_S}(\mbox{WA})&=& 0.41 \pm 0.11. \label{eq:WA-eta'K}
\end{eqnarray}
These should be compared with 
 the world
averaged CP asymmetry determined by the charmonium final states
\begin{eqnarray}
     \sin 2 \phi_1/\beta (\mbox{WA})=0.726 \pm 0.037. \label{eq:sin2phi1}
\end{eqnarray}
Both (\ref{eq:WA-phiK}) and (\ref{eq:WA-eta'K}) appear to differ
from the SM expectation  (\ref{eq:sin2phi1}) significantly by about 2$\sigma$ 
level or so.
Furthermore  the average of CP asymmetries of all $b \to s$ penguin modes
from the Belle results is 
$0.43^{+0.12}_{-0.11}$, being  2.4$\sigma$ away from the SM,
and that of the BaBar results is 
$0.42 \pm 0.10$, corresponding to 2.7$\sigma$ deviation.
Though it may be  premature to conclude  this
to be a clear signal beyond the SM, it is intriguing to note that
the latest results have two characteristic features: 1) the apparent
deviation is seen in more than one mode. 2) Both $S_{\phi
  K^0}(\mbox{WA})$ and $S_{\eta' K_S}(\mbox{WA})$ tend to be
smaller than the SM expectation.

On the theoretical side, study has been done in the literature
\cite{Grossman:1996ke,Barbieri:1997kq,Moroi:2000tk,Lunghi:2001af,Chang:2002mq,Khalil:2002fm,Kane:2002sp,Harnik:2002vs,Ciuchini:2002uv,Agashe:2003rj,Khalil:2003bi,Goto:2003iu,Hisano:2003iw,Chua:2003xq,Mishima:2003ta,Endo:2003te,Cheng:2004jf,Gabrielli:2004yi,Kane:2004ku,Endo:2004fx}
for contributions to $B \to \phi K$ and other modes in the
supersymmetric (SUSY) extension of the SM, a promising candidate for
physics beyond the SM.  An interesting observation was made in
Ref~\cite{Moroi:2000tk,Chang:2002mq,Harnik:2002vs}, pointing out that
large SUSY contribution to $B \to \phi K$ is expected in SUSY seesaw
models embedded into grand-unified theories (GUTs) where
renormalization-group effects above the GUT scale from right-handed
neutrino Yukawa couplings can induce CP-violating flavor mixing in the
mass squared matrix of scalar superpartners of right-handed down-type
quarks. Other possible sources of CP-violating flavor mixing have also
been discussed in the literature
\cite{Agashe:2003rj,Goto:2003iu,Endo:2003te,Kane:2004ku}.  
Ref.~\cite{Endo:2004fx}
has discussed the possibility to discriminate various scenarios of
flavor violation by combining low-energy data accessible at present
and in the future.

In this paper, we would like to discuss implications of the latest
results summarized as (\ref{eq:WA-phiK}) and (\ref{eq:WA-eta'K}) in
the SUSY SM, when we
take the results as an indication of new physics beyond the SM. 
We will stress that the pattern of the deviations from the
SM suggested by the combined results (\ref{eq:WA-phiK}) and 
(\ref{eq:WA-eta'K}) requires a specific structure on the squark mass
matrices, namely new CP violating flavor mixing in the left-handed
squark sector is needed to explain it. As we will discuss later, this
conclusion is rather striking to the model building, especially to
the efforts to explore mediation mechanisms of SUSY breaking, because
renormalization group effects from the Yukawa couplings of quarks and
leptons as well as of right handed neutrinos do not generate this
type of mixing. Thus we are led to invoke some other source to
realize the flavor mixing. This argument disfavors not only the gauge
mediation of the SUSY breaking, but also the minimal supergravity and
some simple class of anomaly mediation. Rather the new CP violating
flavor mixing should be implemented by the mediation mechanism itself,
which  may be based on flavor symmetry or on configuration of quark and
lepton supermultiplets in extra dimensions.

\

We begin with a brief review of the possible new physics contributions.  
The $b \to s$ transition amplitudes are estimated by using the effective 
Hamiltonian,  
\begin{eqnarray}
  H_{\rm eff}\,=\,
  \frac{4G_F}{\sqrt{2}}
  \left[
    \sum_{q'=u,c}V_{q'b}V_{q's}^* 
    \sum_{i=1,2}
    C_i O_i^{(q')}
    - V_{tb}V_{ts}^*
    \sum_{i=3\sim6, 7\gamma, 8G} 
    \left( C_iO_i + \widetilde{C}_i\widetilde{O}_i \right)
  \right]\;,
\end{eqnarray}
where the operators are given by
\begin{eqnarray}
  & &O_1^{(q')}
  = ({\bar s}_i \gamma_\mu P_L q'_j)({\bar q'}_j \gamma^\mu P_L b_i)\;,
  \;\;\;\;\;\;\;\;
  O_2^{(q')}
  = ({\bar s}_i \gamma_\mu P_L q'_i)({\bar q'}_j \gamma^\mu P_L b_j)\;,
  \nonumber \\
  & &O_3
  =({\bar s}_i \gamma_\mu P_L b_i)\sum_{q}({\bar q}_j \gamma^\mu P_L q_j)\;,
  \;\;\;\;\;\;\;
  O_4
  =({\bar s}_i \gamma_\mu P_L b_j)\sum_{q}({\bar q}_j \gamma^\mu P_L q_i)\;, 
  \nonumber \\
  & &O_5
  =({\bar s}_i \gamma_\mu P_L b_i)\sum_{q}({\bar q}_j \gamma^\mu P_R q_j)\;,
  \;\;\;\;\;\;\;
  O_6
  =({\bar s}_i \gamma_\mu P_L b_j)\sum_{q}({\bar q}_j \gamma^\mu P_R q_i)\;, 
  \nonumber \\
  & &O_{7\gamma} 
  =  \frac{e}{16\pi^2} m_b {\bar s}_i \sigma^{\mu\nu}
  P_R b_i F_{\mu\nu}\;,
  \;\;\;\;\;\;\;\;\;
  O_{8G}    
  =  \frac{g_s}{16\pi^2} m_b {\bar s}_i \sigma^{\mu\nu}
  P_R T^a_{ij} b_j G^a_{\mu\nu}\;,
\end{eqnarray} 
with $P_R = (1+\gamma_5)/2$ and $P_L = (1-\gamma_5)/2$.
Here, $i$ and $j$ are color indices, and $q$ is taken to be $u$, $d$, $s$
and $c$.  The terms with tilde are obtained by flipping chiralities, 
$L \leftrightarrow R$. 

In the SUSY SM, there are two kinds of sources of flavor mixings: 
the CKM matrix and squark mass matrices.  
The latter induce large FCNC through the gluino exchange
diagrams.  The off-diagonal elements in the squark mass matrices are
parameterized as~\cite{Hall:1985dx,Gabbiani:1996hi}, 
\begin{eqnarray}
  (\delta^d_{LL})_{23} = \frac{(m^2_{\tilde d_{L}})_{23}}{m^2_{\tilde q}}
  \;,&~~~&
  (\delta^d_{RR})_{23} = \frac{(m^2_{\tilde d_{R}})_{23}}{m^2_{\tilde q}}
  \;, \nonumber\\
  (\delta^d_{LR})_{23} = \frac{(m^2_{\tilde d_{LR}})_{23}}{m^2_{\tilde q}}
  \;,&~~~&
  (\delta^d_{RL})_{23} = (\delta^d_{LR})_{32}^*
  \;,
\end{eqnarray}
where $m^2_{\tilde d}$ is the squared down-type-squark mass matrix,
$m_{\tilde q}$ an averaged squark mass. As we will see shortly, it is 
convenient to divide these mixing parameters into two classes in 
terms of chiralities, which are represented by subscript, $L$ and $R$.  
One class is of the left-handed squark mixings, including $(\delta_{LL}^d)_{23}$ 
and $(\delta_{LR}^d)_{23}$, whereas $(\delta_{RR}^d)_{23}$ and 
$(\delta_{RL}^d)_{23}$ belong to the right-handed class.  
Note that these mixings generally have CP violating phases and thus 
induce CP asymmetries in various observables.

Although the $b \to s$ transition channels are favorite ones to seek for new 
physics as there is no tree-level SM contribution,
 the heaviness of
the superparticles required by null results of superparticles searches
suppresses the SUSY contribution in general. This is particularly the
case for the SUSY contribution to the Wilson coefficients $C_{1-6}$
and $\tilde C_{1-6}$ evaluated at the superparticle mass scale: they
are known to be small even when the flavor violating parameters
$(\delta_{LL}^d)_{23}$ and $(\delta_{RR}^d)_{23}$ are of order 0.1.
On the other hand, the magnetic penguin
 contributions can be quite sizable when the enhancement works for 
large $\tan \beta$, the ratio of the two Higgs vacuum expectation values,
or when the flavor mixing in the left-right
mixing, namely $(\delta_{LR}^d)_{23}$ or $(\delta_{RL}^d)_{23}$, is
not suppressed by the b-quark mass relative to a typical superparticle
mass scale. In the following, we therefore concentrate on the dipole 
type operators when discussing the SUSY contribution.

The gluino contributions to the relevant Wilson coefficients $C_{7\gamma}$
and $C_{8G}$ 
at supersymmetry scale $M_S$ are evaluated to be 
\begin{eqnarray}
  C_{7\gamma}^{\tilde g} (M_S)
  &\simeq&
  -\frac{\sqrt{2} \alpha_s \pi}
  {6G_F V_{tb} V_{ts}^*  m_{\tilde q}^2}
  \left[
    (\delta_{LL}^d)_{23}\, 
    \left(
      \frac{8}{3} M_3(x)
      - 
      \mu_H \tan\beta \frac{m_{\tilde g}}{m_{\tilde q}^2}
      \frac{8}{3} M_a(x)
    \right)
  \right.\nonumber\\
  & & \left.\ \ \ \ \ \ \ \ \ \ \ \ \ \ \ \ \ \ \ 
    +
    (\delta_{LR}^d)_{23}\, \frac{m_{\tilde g}}{m_b}\,
    \frac{8}{3} M_1(x) \right]
  \;,\nonumber\\
  C_{8G}^{\tilde g} (M_S)
  &\simeq&
  -\frac{\sqrt{2} \alpha_s \pi}
  {2G_F V_{tb} V_{ts}^*  m_{\tilde q}^2}
  \left[
    (\delta_{LL}^d)_{23}
    \left\{
      \left( \frac{1}{3} M_3(x) + 3 M_4(x)\right)
    \right.\right.\nonumber\\
  & &\left.\left. \hspace{-5mm}
      - 
      \mu_H \tan\beta \frac{m_{\tilde g}}{m_{\tilde q}^2}
      \left(\frac{1}{3} M_a(x)+ 3M_b(x)\right)
    \right\}
    +\,
    (\delta_{LR}^d)_{23}\, \frac{m_{\tilde{g}}}{m_b}
    \left(\frac{1}{3} M_1(x) + 3 M_2(x)\right)\right]\;.  \label{eq:Wilson}
\end{eqnarray}
Here $M_{1-4}$ and $M_{a,b}$ are the loop functions given in 
Ref.~\cite{Endo:2004fx} and $x = m^2_{\tilde{g}}/m^2_{\tilde{q}}$, where 
$m_{\tilde g}$ is the gluino mass.  The superscript $\tilde g$ represents
gluino-loop contribution.
The Wilson coefficients corresponding to the magnetic type operators 
have the contributions from double mass insertion diagrams, which are 
proportional to  $\tan\beta$ and thus are enhanced if 
$\tan\beta$ is large.  
It is important that the Wilson coefficients $C_i^{\tilde g}$'s 
depend only on the left-handed squark mixings.  On the other hand, 
the tilded $\widetilde{C}_i^{\tilde g}$'s are proportional to the 
right-handed ones.  
Since the left- and right-handed squark mixings generally have an 
independent phase of that of the SM, the gluino contributions may 
induce large CP asymmetry in $b \to s$ transition modes.  

The other source of flavor violations is the CKM matrix.  
This then induces the Wilson coefficients through the diagrams 
mediated by the particles of the SM, the charged Higgs, the 
chargino and the neutralino.  In the following analysis, we estimate 
the SM contributions at the one loop order and partially at the two 
loop level~\cite{Buchalla:1995vs}.  The diagrams of the charged Higgs 
are calculated at the two loop order~\cite{Ciuchini:1997xe}.  
The other SUSY contributions are estimated at the one loop level, including 
the corrections by large $\tan\beta$~\cite{Degrassi:2000qf,Borzumati:2003rr}.  
Here note that all these contributions originate in the CKM matrix.  
Thus they have no additional phase, that is, their phase aligns to that 
of the SM.  Hereafter, we set the phase to be zero after rephasing, 
and the imaginary part of the Wilson coefficients induce the CP asymmetry.  

The mixing-induced CP asymmetry of a CP eigenstate $f_{\rm CP}$ is given by
\begin{eqnarray}
S_{f_{\rm CP}}\,
=\,
\xi_{f_{\rm CP}}
\frac
{2\; {\rm Im}\left[
e^{-2i\,\phi_1} A(\overline{B}_d\to f_{\rm CP})/A(B_d\to f_{\rm CP})
 \right]}
{|A(\overline{B}_d\to f_{\rm CP})/A(B_d\to f_{\rm CP})|^2+1}
\;,
\end{eqnarray}
where $\xi_{f_{\rm CP}}$ is the CP phase of the final state $f_{\rm
  CP}$.  In the SM, the mixing-induced CP asymmetry for the final
states $\phi K_{S(L)}$ and $\eta' K_{S(L)}$ is equal to $\sin
2\phi_1/\beta$ in a good approximation. 
They remain the same if the new physics contributions have no extra 
CP-violating phases.  
New CP-violating phases, on the other hand, 
drive $S_{\phi K^0}$
and/or $S_{\eta' K_S}$ away from the SM prediction.  In the
supersymmetry models the additional phase can  appear in the squark
mixings, giving rise to  the deviation of $S_{\phi K^0}$ and/or
$S_{\eta' K_S}$.

A simple but crucial observation for our argument is that 
the modes $\phi K$ and $\eta' K$ receive different supersymmetric 
contributions because of the chirality structure of local 
operators~\cite{Kagan,Khalil:2003bi}. 
The matrix elements of $O_i$ and $\widetilde{O}_i$ satisfy
\begin{eqnarray}
\langle f | O_i | B_d \rangle
&=&
- (-1)^{P_f}
\langle f | \widetilde{O}_i | B_d \rangle
\,
\end{eqnarray}
where $P_f$ is parity of the final state $f$.  Therefore the
supersymmetric contributions are different between the decay amplitudes
of $\phi K$ and $\eta' K$:
\begin{eqnarray}
A_i^{\tilde g}(B_d\to \phi K)
\,
\propto\,
C_i^{\tilde g}(m_b)\, +\, \widetilde{C}_i^{\tilde g}(m_b)
\;,\\
A_i^{\tilde g}(B_d\to \eta' K)
\,
\propto\,
C_i^{\tilde g}(m_b)\, -\, \widetilde{C}_i^{\tilde g}(m_b)
\;,
\end{eqnarray}
due to parity difference between  $\phi K$ and $\eta'K $ produced by 
$B_d$ meson decay. Consequently, given the pattern of the deviations of  
$S_{\phi K^0}$ and  $S_{\eta' K_S}$, one can infer what types of new physics
contributions are required. 
The present experimental results favor both $S_{\phi K^0}$ and $S_{\eta' K_S}$ 
smaller  than the SM prediction of $\sin
2\phi_1/\beta$, suggesting that the left-handed squark mixing should dominate 
in the  $b$ to
$s$ penguin.   

In $S_{\phi K^0}$ and $S_{\eta' K_S}$, the dominant supersymmetric
contribution comes from the chromo-magnetic penguin. 
Here let us use the generalized factorization method~\cite{Ali:1997nh}
in the calculation of hadronic part.
In this method the main theoretical error comes from the matrix element of
the chromo-magnetic penguin,
\begin{eqnarray}
\langle \phi(\eta') K  | O_{8G} | \overline{B}_d \rangle\,
=\,
-
\frac{\alpha_s(m_b)}{4\pi}\,
\frac{m_b}{\sqrt{q^2}}
\left\langle \phi(\eta') K \left|
O_4 + O_6 -\frac{1}{3}(O_3 + O_5)
\right| \overline{B}_d \right\rangle
\;,
\end{eqnarray}
where $q^2$ is the momentum transferred by the gluon in $O_{8G}$.  
A simple kinematic consideration leads the range of $q^2$ to be
typically $m_b^2/4 \lesssim q^2 \lesssim m_b^2/2$. 
In this study we take  
$q^2=(M_B^2-M_{\phi(\eta')}^2/2+M_K^2)/2$~\cite{Ali:1997nh}. 
Though the
parameter $q^2$ is ambiguous in the generalized factorization, this is
not an inherent uncertainty in the estimation of $S_{\phi K^0 (\eta' K_S)}$.  
In fact this can be removed once we apply the QCD 
factorization~\cite{Beneke:1999br} or the perturbative
QCD~\cite{Keum:2000ph,Mishima:2003wm}.   
However, the fact that $\phi K$ and $\eta' K$ receive different contribution
due to chirality structure is quite generic in all methods, therefore
each method gives similar result as discussed below.   

\begin{figure}[h]
  \begin{center}
    \includegraphics{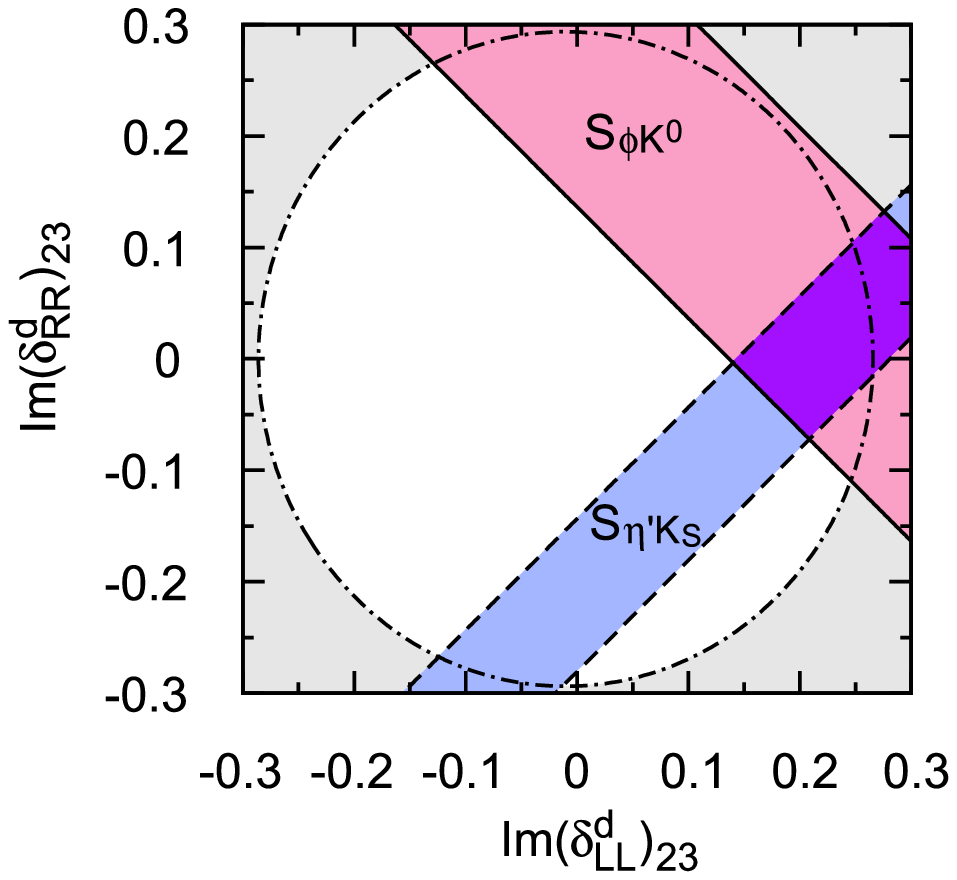}
  \end{center}
  \caption{
    Regions which reproduce the current experimental values of $S_{\phi K^0}$ 
    and $S_{\eta' K_S}$ within 1$\sigma$ errors. The horizontal 
    (vertical) axis is the imaginary part of the left-handed (right-handed) 
    flavor mixing $(\delta^d_{LL})_{23}$  ($(\delta^d_{RR})_{23}$).  The shadowed 
    region outside the dot-dashed line is excluded by ${\rm Br}(b \to s 
    \gamma)$.  The soft SUSY breaking masses as well as the higgsino-mixing 
    mass parameter are taken to be 500 GeV and $\tan\beta = 10$.}
  \label{fig:phiKS_etaKs}
\end{figure}

We show the numerical result of these processes, which have different
dependence of chiralities.  In Fig.~\ref{fig:phiKS_etaKs}, the regions
which reproduce the current experimental values of $S_{\phi K^0}$ and
$S_{\eta' K_S}$ are shown, respectively. Here 1$\sigma$ errors are
considered. The horizontal axis represents the imaginary part of the
left-handed flavor mixing $(\delta^d_{LL})_{33}$, whereas the vertical
axis is its right-handed counter part. Here we take the squark mixings
simply to be pure imaginary and set the left-right mixing to zero,
$(\delta^d_{LR})_{23}=(\delta^d_{RL})_{23}=0$ for simplicity.  All
soft SUSY breaking masses as well as $\mu_H$ at the SUSY scale are
taken to be 500 GeV and $\tan\beta = 10$.  Too large squark mixings
are excluded by the branching ratio of the inclusive decay of $b \to s
\gamma$. Here we conservatively allow rather a broad range, $2.0
\times 10^{-4} < {\rm Br}(b \to s \gamma) < 4.5 \times 10^{-4}$. 
The region where the two bands (one from $S_{\phi K^0}$ and the other from
$S_{\eta' K_S}$)  overlap inside the circle allowed by $b \to s \gamma$
is favored. 
Clearly we find that the present data from the
$B$-factories prefers {\it the left-handed squark mixings} to
the right-handed ones and implies the mixing takes a value of 
$O(0.1)$.\footnote{Inclusion of (or replacement with) the left-right 
mixing does not alter this conclusion.}

This result is very generic.  First, we consider the theoretical uncertainties 
in the estimation of $S_{\phi K^0}$ and $S_{\eta' K_S}$.  As mentioned above, 
the main source originates in the momentum transfer $q^2$.  When we take a 
different value, the squark mixings required to realize the current 
experimental results indeed have to be scaled, but toward the same direction 
for both $S_{\phi K^0}$ and $S_{\eta' K_S}$.  Hence the conclusion remains 
the same, that is, the left-handed mixing should be the dominant source.  
On the other hand, there are other methods for calculating the processes, 
including the QCD factorization.  We expect 
that they will provide a similar conclusion.  In fact, the QCD factorization, 
for example, is found to show a good agreement in the behaviors 
of the modes of $S_{\phi K^0}$ and $S_{\eta' K_S}$ with the naive 
factorization~\cite{Gabrielli:2004yi}.  
Second, there are ambiguities in the contributions from the chargino, 
the neutralino and the charged Higgs, in fact, by choosing a different 
set of the soft parameters.  Notice that the source of the flavor mixings 
in them are based on the CKM matrix, and thus they contribute only to the real 
components of the Wilson coefficients.  In this letter, in contrast, 
we focus on the CP asymmetry, $S_{\phi K^0}$ and $S_{\eta' K_S}$, which turn out 
to be less sensitive to the real components.  
Rather they just modify the allowed ranges of the imaginary part 
of the squark mixings, which is now constrained by 
${\rm Br}(b \to s \gamma)$.  Therefore we draw the same conclusion.  
Third, the phases of the squark mixings are undetermined yet.  
In the above analysis, we simply set them to be maximal. If we instead take 
smaller phases, the imaginary parts become smaller.  
 Hence the current data of $S_{\phi K^0}$ and $S_{\eta' K_S}$ 
require larger magnitude of the squark mixings, leading just to scale 
the result in Fig.~\ref{fig:phiKS_etaKs}.  At the same time, they also 
contribute to the real components of the Wilson coefficients.  
However the CP asymmetric modes depend weakly on them.  As a result, 
the left-handed squark mixing is favored again.  
Finally, we would like to comment on the SM contribution to the $\eta' K_s$ 
mode.  
It is known that the standard computation of the branching  
ratio of the $B_d \to \eta' K_s$ process based on the SM is smaller than what 
is measured. This implies that  there might be some additional 
contributions to the amplitude within the SM, like contribution from gluonium 
and 
so on~\footnote{
  The SUSY contributions to the branching ratio are also discussed in 
  Ref.~\cite{Cheng:2004jf,Gabrielli:2004yi}.  }.  
Since they have no independent phase, they modify only the real components 
in the $\eta' K_s$ mode.  Therefore we expect $S_{\eta' K_S}$ is not affected 
so strongly and the scenario of the dominant left-handed squark mixings will 
go well.  

From Fig.~\ref{fig:phiKS_etaKs}, we should note that pure left-handed
mixing without right-handed one is consistent with the current experimental
results.  Here we would like to stress that this mixing pattern is favored to
avoid the severe constraint on the chromo-electric dipole moment (CEDM)
of the strange quark. Some time ago, Hisano and Shimizu recognized that
 if both left-handed and right-handed 2-3 generation mixings of down-type 
squark masses exist with non-zero relative CP phase, the CEDM of the strange
quark is generated through gluino loop, which is severely constrained by
the measurements of the electric dipole moments (EDMs) of the 
atom~\cite{Hisano:2003iw,Hisano:2004tf} and neutron~\cite{Hisano:2004tf}.
If one notices that the left-handed mixing is generically generated by
the renormalization group effect of the down-type quark Yukawa couplings,
one finds that the existence of the CP violating right-handed squark mixing
would face serious conflict with the CEDM 
constraint~\cite{Hisano:2003iw,Kane:2004ku}.
On the other hand, the present scenario without the right-handed mixing does
not have such a difficulty. In fact, the CP violating left-handed squark mixing
also induces the strange-quark CEDM by chargino exchange, but the constraint 
from this turns out to be much less stringent~\cite{Endo:2003te}.

\

In this paper, we have discussed how to explain the possible anomalies
in the $b \to s$ penguin processes recently measured at the BaBar and
Belle experiments, in the context of supersymmetry. We have found that
the left-handed squark mixings, not the right-handed ones, should be
responsible to account for the present central values of $B \to \phi
K^0$ and $B \to \eta' K_S$ simultaneously. Furthermore the constraint on 
the CEDM of the strange quark favors the pure left-handed
mixings and absence of the right-handed mixings. 

How can we realize such a mixing pattern?  Interestingly it is
implausible that renormalization group effects can generate this
pattern of the squark mixings.  The Yukawa couplings of quarks and
leptons which appear in the minimal supersymmetric standard model
(MSSM) can in fact yield left-handed squark mixings but without new
CP phases, because all CP phase in the MSSM Yukawa coupling comes
solely from the Kobayashi-Maskawa phase. Thus this does not contribute
new CP asymmetry in the $b \to s$ penguin modes. Embedded into GUTs,
the Yukawa couplings can in general contain new CP phases because
under GUTs quarks and leptons in the same GUT multiplets cannot be
rotated freely and so some complex phases remain physical. An
inspection shows, however, such CP phases do not arise in the
left-handed squark mixings. When one considers seesaw mechanism under
SUSY GUTs framework, the right-handed neutrino Yukawa couplings can
generate squark mixings due to the renormalization-group flow above
the GUT scale. However this gives mixings solely in the right-handed
down-type squarks. Thus these well-motivated Yukawa couplings we
mentioned above cannot produce the left-handed squark mixings with
non-trivial CP phases~\footnote{By the same reason, the LR mixing is not 
induced by the renormalization group effects of the Yukawa couplings.}.

This argument should be rather striking to the long-standing efforts
to seek for successful scenarios of mediation of SUSY breaking.  In
fact most of the efforts have been made to realize flavor-blind
SUSY-breaking masses of squarks and sleptons. In this case only source
of flavor violation is in radiative corrections. However, as we argued
just above, this would not generate the CP violating flavor mixing in
the left-handed squark sector. 

Thus we are led to consider the case where the flavor mixing is
imprinted from the beginning, namely, at the very high energy scale
where the SUSY breaking is mediated to the MSSM sector. To avoid too
large FCNC from squark loops, the squark mass matrices should be
aligned with their quark counterparts at least in the first two
generations, which can be realized with the help of flavor symmetry or
configuration of quark and lepton supermultiplets in extra dimensions.
In this alignment mechanism, there exist in general CP-violating
squark mixings. Now the question is how to suppress the right-handed
ones. We feel that this is quite a non-trivial requirement on the
model building. A possible way to achieve this will be to invoke
localizations of fields along the extra dimensions. In fact suppose
that by some reason each generation of the right-handed quark
supermultiplets is localized at a different point of extra dimensions
and the overlap of the wave functions between the fields in the
different generations is sufficiently suppressed. Then the generation
mixing in the right-handed squark masses should also be
suppressed. See, {\it e.g.} \cite{Watari:2002fd}, which considered
localization of $\bar 5$ multiplets in SU(5) GUT at orbifold fixed
points in a different context.

Before closing, we would like to briefly mention implications of our
scenario to other observables.  The result of the large left-handed
squark mixings and the suppressed RR one leads to the following
result: the direct CP asymmetry of the inclusive $b \to s \gamma$,
$A_{\rm CP}(b \to s \gamma)$, can be as large as a few percent with a
positive sign, which can be detectable at $B$-factories and super
$B$-factory.  The mixing-induced CP asymmetry of $B \to K^* \gamma$,
$S_{K^* \gamma}$, is sensitive to the $\tilde C_{7\gamma}$ and thus
the present result of $S_{\phi K^0}$ and $S_{\eta' K_S}$ implies the
suppression of $S_{K^* \gamma}$.  Also the mass difference in the
$B_s-\overline{B_s}$ mixing, $\Delta M_s$, is enhanced only when both
the LL and RR squark mixings are large, hence the deviation of $\Delta
M_s$ from the SM will be small.  By measuring $S_{\phi K^0}$,
$S_{\eta' K_S}$ more accurately and combing them with these additional 
quantities which may be observed at $B$-factories and super
$B$-factory, we will be able to test whether  the left-handed squark mixing
is really the dominant source of the flavor mixings.

\section*{Acknowledgment}
M.E. thanks the Japan Society for the Promotion of Science for financial 
support.
This work  was supported in part by the Grants-in-aid from the Ministry
of Education, Culture, Sports, Science and Technology, Japan, No.14046201 and
No.16081202.

\clearpage

\clearpage

\begin{thebibliography}{99}
\bibitem{Belle}
Y.~Sakai (Belle collaboration), talk given at the 32nd International Conference
on High Energy Physics (ICHEP'04), August 16-22, 2004, Beijing, China. 
http://ichep04.ihep.ac.cn; 
K.~Abe {\it et al.} [the Belle Collaboration], hep-ex/0409049.

\bibitem{BaBar}
M.~Giorgi (BaBar collaboration), talk given at the 32nd International Conference
on High Energy Physics (ICHEP'04), August 16-22, 2004, Beijing, China. 
http://ichep04.ihep.ac.cn; 
B.~Aubert {\it et al.}  [BABAR Collaboration],
arXiv:hep-ex/0408072;
arXiv:hep-ex/0408090.


\bibitem{Ligeti}
Z. Ligeti, talk given at  the 32nd International Conference
on High Energy Physics (ICHEP'04), August 16-22, 2004, Beijing, China. 
http://ichep04.ihep.ac.cn;
arXiv:hep-ph/0408267.

\bibitem{Grossman:1996ke}   
Y.~Grossman and M.~P.~Worah,   
Phys.\ Lett.\ B {\bf 395}, 241 (1997).
   
\bibitem{Barbieri:1997kq}   
R.~Barbieri and A.~Strumia,   
Nucl.\ Phys.\ B {\bf 508} 3 (1997).
   
\bibitem{Moroi:2000tk}   
T.~Moroi,   
Phys.\ Lett.\ B {\bf 493}, 366 (2000).
   
\bibitem{Lunghi:2001af}   
E.~Lunghi and D.~Wyler,   
Phys.\ Lett.\ B {\bf 521} 320 (2001). 
   
   
   
\bibitem{Chang:2002mq}   
D.~Chang, A.~Masiero and H.~Murayama,   
Phys.\ Rev.\ D {\bf 67}, 075013 (2003).
   
\bibitem{Khalil:2002fm}   
S.~Khalil and E.~Kou,   
Phys.\ Rev.\ D {\bf 67}, 055009 (2003).
   

   
\bibitem{Kane:2002sp}   
G.~L.~Kane, P.~Ko, H.~b.~Wang, C.~Kolda, J.~H.~Park and L.~T.~Wang,   
Phys.\ Rev.\ D {\bf 70}, 035015 (2004);
Phys.\ Rev.\ Lett.\  {\bf 90}, 141803 (2003).
   
\bibitem{Harnik:2002vs}
R.~Harnik, D.~T.~Larson, H.~Murayama and A.~Pierce,
Phys.\ Rev.\ D {\bf 69}, 094024 (2004).

\bibitem{Ciuchini:2002uv}   
M.~Ciuchini, E.~Franco, A.~Masiero and L.~Silvestrini,   
Phys.\ Rev.\ D {\bf 67}, 075016 (2003)   
[Erratum-ibid.\ D {\bf 68}, 079901 (2003)].
   
\bibitem{Agashe:2003rj}   
K.~Agashe and C.~D.~Carone,   
Phys.\ Rev.\ D {\bf 68}, 035017 (2003).
   

 
 
\bibitem{Khalil:2003bi}
S.~Khalil and E.~Kou,
Phys.\ Rev.\ Lett.\  {\bf 91}, 241602 (2003).

\bibitem{Goto:2003iu}
T.~Goto, Y.~Okada, Y.~Shimizu, T.~Shindou and M.~Tanaka,
Phys.\ Rev.\ D {\bf 70}, 035012 (2004).


  
\bibitem{Hisano:2003iw}   
J.~Hisano and Y.~Shimizu,   
Phys.\ Lett.\ B {\bf 581}, 224 (2004).
   
\bibitem{Chua:2003xq}   
C.~K.~Chua, W.~S.~Hou and M.~Nagashima,   
Phys.\ Rev.\ Lett.\  {\bf 92}, 201803 (2004).
   
   
   
\bibitem{Mishima:2003ta}   
S.~Mishima and A.~I.~Sanda,   
Phys.\ Rev.\ D {\bf 69}, 054005 (2004).
   
\bibitem{Endo:2003te}   
M.~Endo, M.~Kakizaki and M.~Yamaguchi,   
Phys.\ Lett.\ B {\bf 583}, 186 (2004);
arXiv:hep-ph/0403260.   
   
\bibitem{Cheng:2004jf}
J.~F.~Cheng, C.~S.~Huang and X.~H.~Wu,
arXiv:hep-ph/0404055.

   
\bibitem{Gabrielli:2004yi}   
E.~Gabrielli, K.~Huitu and S.~Khalil,   
arXiv:hep-ph/0407291.   
   
   
   
\bibitem{Kane:2004ku}   
G.~L.~Kane, H.~b.~Wang, L.~T.~Wang and T.~T.~Wang,   
arXiv:hep-ph/0407351.   




\bibitem{Endo:2004fx}
M.~Endo and S.~Mishima,
arXiv:hep-ph/0408138.


\bibitem{Hall:1985dx}
L.~J.~Hall, V.~A.~Kostelecky and S.~Raby,
Nucl.\ Phys.\ B {\bf 267}, 415 (1986).

\bibitem{Gabbiani:1996hi}
F.~Gabbiani, E.~Gabrielli, A.~Masiero and L.~Silvestrini,
Nucl.\ Phys.\ B {\bf 477}, 321 (1996).

\bibitem{Buchalla:1995vs}
G.~Buchalla, A.~J.~Buras and M.~E.~Lautenbacher,
Rev.\ Mod.\ Phys.\  {\bf 68} 1125 (1996).

\bibitem{Ciuchini:1997xe}
M.~Ciuchini, G.~Degrassi, P.~Gambino and G.~F.~Giudice,
Nucl.\ Phys.\ B {\bf 527} 21 (1998);\\
F.~M.~Borzumati and C.~Greub,
Phys.\ Rev.\ D {\bf 58} 074004 (1998).

\bibitem{Degrassi:2000qf}
G.~Degrassi, P.~Gambino and G.~F.~Giudice,
JHEP {\bf 0012} 009 (2000);\\
M.~Carena, D.~Garcia, U.~Nierste and C.~E.~M.~Wagner,
Phys.\ Lett.\ B {\bf 499} 141 (2001).

\bibitem{Borzumati:2003rr}
F.~Borzumati, C.~Greub and Y.~Yamada,
Phys.\ Rev.\ D {\bf 69} 055005 (2004).

\bibitem{Kagan}
A.~L.~Kagan, lecture at the 30th SLAC Summer Institute on Particle 
Physics: Secrets of the B Meson, 5-16 August 2002, 
Stanford, USA. http://www-conf.slac.stanford.edu/ssi/2002/
 
\bibitem{Ali:1997nh}
A.~Ali and C.~Greub,
Phys.\ Rev.\ D {\bf 57}, 2996 (1998);
%
A.~Ali, G.~Kramer and C.~D.~Lu,
Phys.\ Rev.\ D {\bf 58}, 094009 (1998).

\bibitem{Beneke:1999br}
M.~Beneke, G.~Buchalla, M.~Neubert and C.~T.~Sachrajda,
Phys.\ Rev.\ Lett.\  {\bf 83}, 1914 (1999);
%
        Nucl.\ Phys.\ B {\bf 591}, 313 (2000).

\bibitem{Keum:2000ph}
Y.~Y.~Keum, H.~n.~Li and A.~I.~Sanda,
Phys.\ Lett.\ B {\bf 504}, 6 (2001);
%
Phys.\ Rev.\ D {\bf 63}, 054008 (2001).

\bibitem{Mishima:2003wm}
S.~Mishima and A.~I.~Sanda,
Prog.\ Theor.\ Phys.\  {\bf 110}, 549 (2003)


\bibitem{Hisano:2004tf}
J.~Hisano and Y.~Shimizu,
arXiv:hep-ph/0406091.

\bibitem{Watari:2002fd}
T.~Watari and T.~Yanagida,
Phys.\ Lett.\ B {\bf 544}, 167 (2002).




\end{thebibliography}
\end{document}